\documentclass[aps,prl,,twocolumn,showpacs,showkey,square,numbers,amssymb,amsmath,nobibnotes,nofootinbib,superscriptaddress]{revtex4}
\usepackage{bm}
\usepackage{amsmath}
\usepackage{amsfonts}
\usepackage{amssymb}
\usepackage{eucal}
\usepackage{verbatim}
\usepackage{times}
\usepackage{graphicx}

\def\Ord{\mathcal{O}}
\newcommand{\barr}{\begin{eqnarray}}
\newcommand{\earr}{\end{eqnarray}}
\newcommand{\avg}[1]{\left< #1 \right>}

\begin{document}

%\preprint{}

\title{The Index Distribution of Gaussian Random Matrices}



\author{Satya N. Majumdar}
\author{C\'{e}line Nadal}
\affiliation{Laboratoire de Physique Th\'{e}orique et Mod\`{e}les
Statistiques (UMR 8626 du CNRS), Universit\'{e} Paris-Sud,
B\^{a}timent 100, 91405 Orsay Cedex, France}

\author{Antonello Scardicchio} 
\author{Pierpaolo Vivo}
\affiliation{Abdus Salam International Centre for
Theoretical Physics, Strada Costiera 11, 34014 Trieste, Italy}

\date{\today}

\begin{abstract}
We compute analytically, for large $N$, the probability distribution
of the number of positive eigenvalues (the index $N_{+}$) of a 
random 
$N\times N$ matrix 
belonging to Gaussian orthogonal $(\beta=1)$, unitary $(\beta=2)$ or
symplectic $(\beta=4)$ ensembles. The distribution of the
fraction of positive eigenvalues $c=N_{+}/N$ scales, for large $N$, 
as 
$\mathcal{P}(c,N)\simeq\exp\left[-\beta N^2 \Phi(c)\right]$ 
where the rate function $\Phi(c)$, symmetric around $c=1/2$ and
universal (independent of $\beta$), is calculated 
exactly. The distribution has non-Gaussian tails, but even near its peak
at $c=1/2$ it is not strictly Gaussian due to an unusual logarithmic 
singularity in the rate function.

\end{abstract}



\maketitle

Random matrix theory (RMT) has played a central role in various branches
of physics 
since its inception~\cite{Mehta}. Through the years, different, seemingly 
unrelated problems in physics and mathematics have been linked via RMT. 
It is not surprising then that the distributions of observables associated
with random matrices play a very important role in a variety
of physical contexts.  
Still, after more than half a century, certain natural 
questions about eigenvalue distributions 
have eluded a thorough treatment, in spite of their relevance 
to a broad range of subjects. 

As an example, classical disordered systems offer the ideal environment where 
RMT ideas and tools may be applied. Physical systems such as liquids and
spin glasses are known to exhibit a rich energy or free energy landscape 
characterized by many extrema (minima, maxima and saddles) and rather
complex stability patterns~\cite{Wales} which play an important
role both in statics and dynamics of such systems. The stability of a stationary
point of an $N$-dimensional potential landscape $V(x_1,x_2,\ldots, x_N)$
is decided by the $N$ real eigenvalues of the Hessian matrix $H_{ij}=[\partial^2 
V/{\partial x_i \partial x_j}]$ which is evidently symmetric. If all $N$ 
eigenvalues are positive 
(negative), the stationary point is a local minimum (local maximum). If some, but 
not all, are positive it is a saddle. The number of positive eigenvalues
$0\le N_{+}\le N$, called the index, is a key object of interest as it 
determines the
number of directions in which a stationary point is stable.

In many situations, important insights about the system can be gained by simply 
assuming that the
Hessian is a real symmetric random matrix drawn from a Gaussian ensemble:
Gaussian orthogonal ensemble characterized by the Dyson index $\beta=1$. 
This
{\em random Hessian model} (RHM) has been studied extensively
in the context of disordered systems~\cite{CGG}, landscape based string 
theory~\cite{Douglas,aazami} and quantum 
cosmology~\cite{Mersini-Houghton05}. In RHM, the fraction of
positive eigenvalues $c=N_{+}/N$ is a random variable whose
distribution ${\mathcal P}(c,N)$ is our main object of interest
in this Letter. Although in RHM $\beta=1$, it is also of interest
to study the index distribution for other
Gaussian ensembles namely
the unitary ($\beta=2$) and the symplectic ($\beta=4$). In this Letter,
we study the index distribution for general $\beta>0$.

Due to the Gaussian symmetry of the ensemble, it is clear that
on average half of the eigenvalues are positive (or negative), implying
$\langle c\rangle=1/2$. Thus, the distribution ${\mathcal P}(c,N)={\mathcal 
P}(1-c,N)$ 
must
be symmetric around $c=1/2$ with a peak at $c=1/2$. Cavagna {\it et al.}
studied ${\mathcal P}(c,N)$, for $\beta=1$, using the replica method and 
some additional approximations~\cite{CGG}. They argued 
that
for large $N$ the distribution has a Gaussian peak around its mean~\cite{CGG} 
\begin{equation}
{\mathcal P}(c,N) \approx \exp\left[- \frac{\pi^2 N^2}{2\ln 
(N)}\left(c-1/2\right)^2\right]     
\label{gauss1}
\end{equation}
indicating that the variance $\langle (c-1/2)^2\rangle \approx \ln(N)/{\pi^2 
N^2}$, or equivalently $\langle (N_{+}-N/2)^2\rangle \approx \ln(N)/{ 
\pi^2}$ for large $N$. In the opposite limit, near the tail $c=1$ (or 
equivalently $c=0$)
${\mathcal P}(1,N)$, the probability that all eigenvalues are positive, was 
recently computed exactly for large $N$ and for all $\beta$~\cite{DM}, 
\begin{equation}
{\mathcal P}(1,N) \approx \exp\left[-\beta \theta N^2 \right]; \quad 
\theta=\frac{1}{4}\ln(3).
\label{dm1}
\end{equation}  
It is then natural to ask how does the distribution behave in between, 
i.e., for $0<c<1$.
In particular, the presence of $\ln(N)$ term in (\ref{gauss1}) presents
a challenging puzzle: how does one smoothly interpolate the
distribution between
the two limits, the peak and the tails?

\begin{figure}[htbp]
\includegraphics[height=4.8cm]{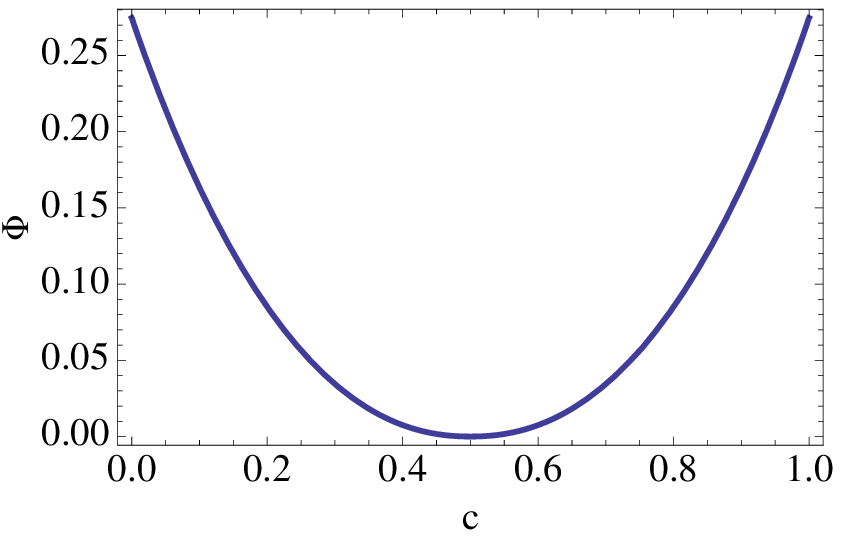}\hspace{0.5cm}
\includegraphics[height=5cm]{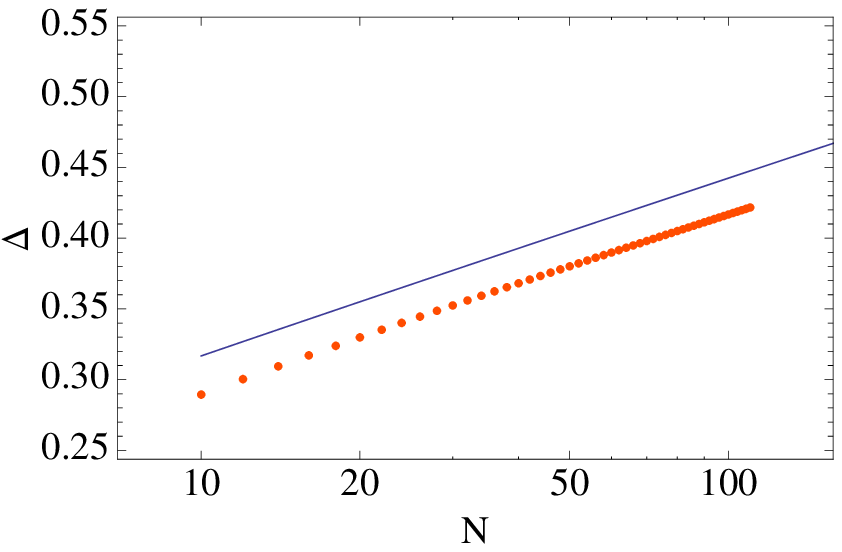}
\caption{(Top) The large deviation function $\Phi(c)$; 
(Bottom) the variance of the index as a function of 
$\ln(N)$ for $\beta=2$ (dotted, exact finite $N$ formula; solid, large $N$). 
A linear fit for the former gives $\Delta(N)\simeq
0.176 + 0.052 
\ln N$ with the prefactor $0.052$ in good agreement with the leading 
theoretical
prefactor $(2\pi^2)^{-1}\simeq 0.051$ for the large $N$ result.} 
\label{fig:phic}
\end{figure}

In this Letter we resolve this outstanding puzzle by computing 
the distribution ${\mathcal P}(c,N)$ exactly for large $N$ in
the full range $0\le c\le 1$ and for all $\beta$. Let us
summarize our main results. We show that to leading order for large $N$,
\begin{equation}
{\mathcal P}(c,N) \approx \exp\left[-\beta N^2 \Phi(c)\right]
\label{ld1}
\end{equation}
where $\approx$ indicates $\lim_{N\to \infty} -\ln\left[{\mathcal 
P}(c,N)\right]/(\beta N^2)= \Phi(c)$. The
exact rate function $\Phi(c)$, symmetric around $c=1/2$ 
and universal (independent of $\beta$), is given in 
(\ref{phic1}) for
$1/2\le c\le 1$ and is plotted in Fig. (\ref{fig:phic}). 
The fact that the logarithm of the probability $\sim O(N^2)$ for fixed $c$
is quite natural, as it represents the free energy of
an associated Coloumb fluid of $N$ charges (eigenvalues).
The Coulomb energy of $N$ charges clearly scales as $\sim O(N^2)$.
In the
limit $c\to 1$, we get $\Phi(1)=\theta= \ln(3)/4$ in agreement with
(\ref{dm1}). The distribution is thus highly non-Gaussian near
its tails. In the opposite 
limit $c\to 1/2$, we find a marginally quadratic
behavior, modulated by an unusual logarithmic singularity
\begin{equation}
\label{gausslog1}
\Phi(c)\simeq -\frac{\pi^2}{2}\frac{(c-1/2)^2}{\ln(c-1/2)}.
\end{equation} 
The variance computed from our exact formula, $\langle (N_{+}-N/2)^2\rangle 
\approx \ln(N)/{\beta \pi^2}$ for large $N$ perfectly agrees, for $\beta=1$, with 
Ref.~\cite{CGG}. However, the distribution of $N_{+}$ near its peak is not a
Gaussian
with variance $\sim \ln(N)$ as claimed in ~\cite{CGG}, rather our exact
result shows that the origin of the $\ln(N)$ term is due to the 
logarithmic
singularity associated with the rate function $\Phi(c)$ itself near $c=1/2$.
In addition to obtaining the full distribution ${\mathcal P}(c,N)$ thus
solving this challenging puzzle, our Coloumb gas approach also
provides a new method of finding solutions to singular integral
equation with two disjoint supports. This method is rather general
and can be fruitfully applied to other related problems in RMT.

Our starting point is the joint distribution of $N$
eigenvalues of Gaussian random matrices parametrized by the Dyson 
index $\beta$~\cite{Mehta}  
\begin{equation}
P(\lambda_1,\ldots,\lambda_N)=
\frac{1}{Z_N}e^{-\frac{\beta}{2}\sum_{i=1}^N\lambda_i^2}
\prod_{1\leq j<k\leq N}|\lambda_j-\lambda_k|^\beta
\end{equation}
where the normalization constant $Z_N$ can be computed using the 
celebrated Selberg's integral. For large $N$, it is known that~\cite{Mehta},
$Z_N\sim \exp\left[-\beta \Omega_0 N^2\right]$ with $\Omega_0=(3+2\ln 2)/8$.

The distribution $P(N_+,N)$ of the index, i.e., the number of positive
eigenvalues can be expressed in terms of the joint distribution
\begin{equation}
\label{PNplus}
P(N_+,N)=\frac{1}{Z_N}\binom{N}{N_+}
\int_{\mathbb{R}^+} \!\!\!\! d^{\scriptscriptstyle{N_+}}
\lambda\int_{\mathbb{R}^-} 
\!\!\!\!  d^{\scriptscriptstyle{N-N_+}}
\lambda\ e^{-\frac{\beta}{2} E( \{ \lambda_i \})}
\end{equation}
where the integrals are restricted over configurations with only $N_+$ 
positive eigenvalues and the binomial counts the different 
relabellings of the eigenvalues. 
The function $E(\{ \lambda_i \})=\sum_i 
\lambda_i^2-\sum_{j\neq k}\ln|\lambda_j-\lambda_k|$ 
can be interpreted as the energy of a configuration of charged particles 
located at $\{\lambda_i\}$ on the real line
and $P(N_+,N)$ is the partition function of this fluid at inverse temperature 
$\beta/2$. The distribution of the fraction $c=N_{+}/N$ is then simply
${\mathcal P}(c,N)= P(cN,N)$.

In this Coulomb gas picture, $N_+=cN$ of the total $N$ charges are confined to 
the positive real semiaxis. The charges repel each other via 
the 2-d Coulomb interaction (logarithmic) and are also subject
to an external confining potential (parabolic). As a result
of these two competing energy scales, it is easy to
see that typically $\lambda\sim O(\sqrt{N})$ for large $N$~\cite{DM}. 
The evaluation of such partition function in the large $N$ limit
is carried out in two steps~\cite{DM}: i) a coarse-graining 
protocol, where one sums over all microscopic arrangements of $\lambda_i$ compatible 
with a fixed and normalized (to unity) charge density function 
$\rho(\lambda,N)=N^{-1}\sum_i\delta(\lambda-\lambda_i)$, and
ii) a functional integral over all possible normalized charge density 
functions, upon using the scaling $\rho(\lambda,N)\sim N^{-1/2}\rho(\lambda 
N^{-1/2})$ where the scaling function $\rho(x)$ satisfies $\int_{-\infty}^{\infty} 
\rho(x) dx=1$.

The resulting functional integral over $\rho(x)$ is then evaluated in the large 
$N$ limit via a saddle point method.
In physical terms, this amounts to finding the equilibrium density of the fluid 
(minimizing its free energy) under the competing interactions 
(Coulomb repulsion and quadratic confinement)
and the external constraint ($N_+=cN$ particles
kept always on the positive semiaxis). This constrained Coulomb gas
approach has proven useful in a number of different contexts
such as Gaussian and Wishart 
extreme eigenvalues 
\cite{DM,vivo2007large,majumdar:060601}, nonintersecting 
Brownian interfaces \cite{nadal:061117}, 
quantum transport in chaotic cavities \cite{vivo:216809}, 
statistics of critical points in Gaussian landscapes
\cite{bray:150201,fyod}, 
bipartite entanglement \cite{facchi:050502} and 
also in information and communication systems \cite{kaz}.

\begin{figure}[htbp]
\includegraphics[bb = 0 0 240 157, height=4.5cm]{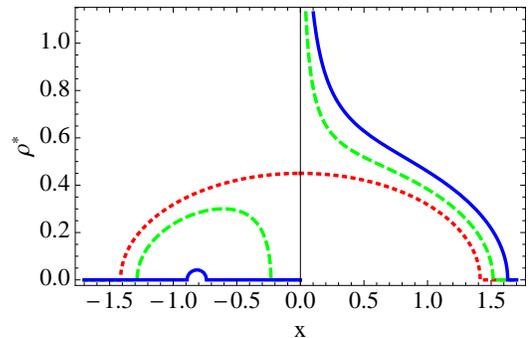}

\caption{The density of eigenvalues $\rho^\star(x)$ (eq. \eqref{sol1}) 
for $c=1/2$ (red), 
$3/4$ (green)  and $0.995$ (blue).}
\label{fig:aLrho}
\end{figure}

Using the above approach one gets, to leading 
order in large $N$,
\begin{equation}\label{decayaction}
\mathcal{P}(c,N)\propto\int\mathcal{D}[\rho]e^{-\frac{\beta}{2} N^2 S_c[\rho]}
\end{equation}
with the action $S_c[\rho]$ given by:
\begin{widetext}
\begin{equation}\label{action}
S_c[\rho]=
\int_{-\infty}^\infty dx\ x^2 \rho(x)-\int_{-\infty}^\infty\int_{-\infty}^\infty 
dx dx^\prime \rho(x) \rho(x^\prime)\ln|x-x^\prime|+
A_1\left(\int_{-\infty}^\infty dx\theta(x)\rho(x)-c\right)+ 
A_2\left(\int_{-\infty}^\infty dx \rho(x)-1\right) 
\end{equation}
\end{widetext}
where $A_1$ and $A_2$ are Lagrange multipliers enforcing the fraction $c$ of 
positive 
eigenvalues and the normalization of $\rho$, and $\theta(x)$
is the Heaviside step function.

The equilibrium fluid density $\rho^\star(x)$, which minimizes the action 
or the free energy, 
is obtained by the saddle point equation $\delta S_c[\rho]/\delta\rho=0$, resulting
in the integral equation
\begin{equation}
x^2 + A_1 \theta(x) + A_2 = 2\int_{-\infty}^{\infty} \rho^\star(y) \ln|x-y|dy
\label{inteq1}
\end{equation}
By taking one derivative with respect of $x$, we obtain, for $x\ne 0$,
\begin{equation}
\label{eq:saddlerho1}
x=\mathrm{Pr}\int dy\frac{\rho^\star(y)}{x-y}.
\end{equation}
where $\mathrm{Pr}$ stands for the Cauchy's principal part. It turns out
that there exists a closed formula, due to Tricomi~\cite{Tricomi}, for the 
solution of such integral
equations provided the solution has a single support.
This is indeed the case in the two limiting situations $c=1/2$ and $c=1$.
For $c=1/2$, the solution is given by the celebrated Wigner's semicircle,
$\rho^\star(x) = \frac{1}{\pi}\sqrt{2-x^2}$ with $-\sqrt{2}\le x\le \sqrt{2}$.
In the opposite limit $c=1$, all the eigenvalues are on the positive side
and one again obtains a single support solution~\cite{DM}, 
$\rho^\star(x)=(2\pi)^{-1}\sqrt{(\sqrt{8/3}-x)/x}\ \left[\sqrt{8/3}+2x\right]$
for $0\le x\le \sqrt{8/3}$. 

\begin{figure}[htbp]
\includegraphics[height=3.6cm]{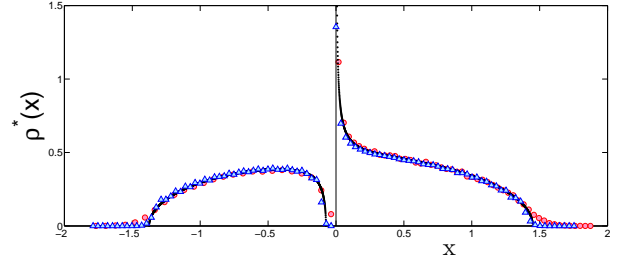}
\caption{Analytical density $\rho^\star(x)$ \eqref{sol1} for $c=0.6$ 
(solid black) together with results from i) (red) numerical diagonalization of 
$10^6$ matrices of size $20\times 20$, where only 
samples having $12$ positive eigenvalues were retained 
for the statistics ($c=0.6$), and ii) (blue) Montecarlo simulations of 
the Coulomb fluid with $N=50$ particles.}
\label{fig:densnum}
\end{figure}

In contrast, for $1/2<c<1$, the solution generally consists of two 
disjoint supports: a blob of $(1-c)N$ negative eigenvalues
and a blob of $cN$ positive eigenvalues (see Fig. \ref{fig:aLrho}).
Finding this two-support solution thus poses the principal technical challenge
for $1/2<c<1$. We have succeeded in finding this two-support solution 
exactly by iterating the Tricomi formula for single-support solution
twice, the details of which will be reported elsewhere~\cite{details}.
Our main result is that for all $1/2\le c\le 1$,  
\begin{equation}
\rho^\star(x)=\frac{1}{\pi} 
\sqrt{\frac{(L-x)}{x}\left(x+L/a\right)\left(x+(1-1/a)\,L\right)}
\label{sol1}
\end{equation}
where $a,L$ parametrize the support of the solution 
$x\in[-L/a,-L(1-1/a)]\cup [0,L]$ where $\rho^\star(x)>0$. 
They are implicitly given as functions of $c$ by the equations:
\begin{equation}
\int_0^1 dy\sqrt{\frac{1-y}{y}}\,\sqrt{y^2+y+ 
\frac{(a-1)}{a^2}}=\frac{\pi c}{2}\,\left(1-\frac{(a-1)}{a^2}\right)
\label{adef}
\end{equation}
and
\begin{equation}
L = \frac{a\sqrt{2}}{\sqrt{a^2-a+1}}.
\label{Ldef}
\end{equation}
It is easy to check that one recovers the correct single support
solutions in the limiting situations $c= 1/2$ and $c=1$. 
The density 
$\rho^\star(x)$ 
for different $c$ is given in Fig. \ref{fig:aLrho}.
We have
also verified this analytical prediction numerically by
two different methods (see Fig. \ref{fig:densnum}): direct diagonalization of 
small matrices
and also by Monte Carlo simulation of the Coloumb gas~\cite{details}.

Using this saddle point solution in
\eqref{decayaction}, we get 
$\mathcal{P}(c,N)\approx\exp(-\beta N^2\Phi(c))$, where 
the rate function $\Phi(c)=(1/2)S_c[\rho^\star]-\Omega_0$,
$\Omega_0=(3+2\ln 2)/8$ coming from the normalization $Z_N$.
To evaluate the saddle point action $S_c[\rho^\star]$  
we next need to evaluate 
the single and double integral over $\rho^\star(x)$ in \eqref{action}. 
The double integral can be written in terms of a simple integral after 
multiplying (\ref{inteq1}) by $\rho^\star$ and then integrating over $x$.
This gives 
\begin{equation}
\Phi(c)=-\frac{3}{8}-\frac{\ln(2)}{4}+\frac{1}{4}\avg{x^2}-
\frac{1}{4}A_1 c-\frac{1}{4}A_2,
\end{equation}
where the average $\langle\cdot\rangle$ is done with the measure $\rho^\star(x)$. 

The Lagrange multipliers$A_1$ and $A_2$ can be obtained  
from (\ref{inteq1}) upon setting $x=L>0$ and $x=-L/a<0$
\begin{eqnarray}
L^2 + A_1 + A_2 &=& 2 \avg{\ln(L-x)}\label{lm1}, \\
L^2/a^2 + A_2 & =& 2 \avg{\ln(x+L/a)}\label{lm2}.
\end{eqnarray}
It turns out that the averages on the right hand side can be
simplified by first introducing a pair of functions
\begin{equation}
W_{(\pm)}(x) = x-\frac{1}{x} \mp 
\sqrt{\frac{(x\mp L)}{x}\left(x\pm\frac{L}{a}\right)
\left(x\pm\left(1-\frac{1}{a}\right)L\right)} 
\label{w12} 
\end{equation}
defined respectively for $x>L$ and $x>L/a$. One can
show~\cite{details} that they are essentially the moment generating
functions of $\rho^*(x)$. The averages
in \eqref{lm1} and \eqref{lm2} can then be expressed as simple integrals
over $W_{(\pm)}(x)$. Skipping details~\cite{details}, we obtain
the following explicit expression for the rate function for $1/2\le c\le 1$,
\begin{widetext}
\begin{equation}
\Phi(c) = \frac{1}{4}[L^2-1-\ln(2L^2)] + \frac{(1-c)}{2}\,\ln(a)
-\frac{(1-c)(a^2-1)}{4a^2}\,L^2
+ \frac{c}{2}\int_{L}^{\infty}W_{(+)}(x) 
dx+\frac{(1-c)}{2}\int_{L/a}^{\infty} W_{(-)}(x)dx.
\label{phic1}
\end{equation}
\end{widetext}

Eq. \eqref{phic1} is the principal result of this Letter. It is again
easy to check that in the two limits $c=1/2$ and $c=1$, one recovers
correctly $\Phi(1/2)=0$ and $\Phi(1)=(\ln 3)/4$.
For arbitrary $c$, the integrals have to be evaluated 
numerically. 
The rate function $\Phi(c)$ is plotted for $0\le c\le 1$ in the top panel 
of Figure \ref{fig:phic}. It is symmetric around $c=1/2$ with a minimum at 
$c=1/2$ and grows monotonically from $\Phi(1/2)=0$ to $\Phi(1)=(\ln 3)/4$.
To see how $\Phi(c)$ behaves near its minimum $c=1/2$, we
make a perturbation expansion of \eqref{phic1} setting $c=1/2+\delta$
with $\delta>0$ small. Since for $c=1/2$, $a=1$, we first expand
\eqref{adef} setting $a=1+\alpha$. To leading order we get
$\alpha\approx \pi\delta/\ln(1/\delta)$. Inserting this result
in \eqref{phic1} followed by straightforward expansion gives
\eqref{gausslog1}. Using this expression of $\Phi(c)$ in
\eqref{ld1}, one can then easily compute the variance of $N_+=cN$
\begin{equation}\label{varianceN}
\Delta(N)=N^2\langle (c-1/2)^2\rangle \simeq\frac{1}{\beta \pi^2}\ln N+\Ord(1)
\end{equation}
which, for $\beta=1$, agrees with the asymptotic result in ~\cite{CGG}.
However, the logarithmic growth of the variance is evidently due
to the logarithmic singularity in the rate function $\Phi(c)$ itself
in \eqref{gausslog1} and the index distribution is strictly not Gaussian
near $c=1/2$.   

We also remark that for $\beta=2$ it is possible to find an exact formula for the 
variance at \emph{finite} $N$ \cite{details} based on Andrejeff formula and/or 
orthogonal polynomials with discontinuous weights
\cite{chen2005orthogonal}. Using this formula we have evaluated
the variance for all finite $N$ and found that the leading
growth is precisely $\Delta(N)\sim \ln(N)/{2 \pi^2}$ in agreement  
with the asymptotic result in \eqref{varianceN}, they 
differ only 
for subleading terms in $N$ (see Fig. \ref{fig:phic} (bottom)).

The work presented here can be generalized in several directions. Our method
to find explicitly two-support solutions to singular integral equation
is quite general and can be applied to other problems. For example,
one can compute the distribution of the number of eigenvalues
bigger than a fixed value $z$. This is particularly relevant
for Wishart matrices that play an important role in 
multivariate data analysis~\cite{majumdar:060601}. Our method
can also be applied to more exotic ensembles of random matrices that 
arise in connection with 2-d quantum gravity  \cite{di19952d}.
It would also be interesting to investigate if the logarithmic 
growth of the variance of the index is universal and holds
even for non-Gaussian ensembles.

%\bibliography{index-letter}

\begin{thebibliography}{99}


\bibitem{Mehta} M.L. Mehta, {\it Random Matrices} (Academic Press, Boston, 1991).



\bibitem{Wales} D.J. Wales, {\it Energy Landscapes: 
Applications to Clusters, Biomolecules and Glasses} (Cambridge University 
Press, 2004).

\bibitem{CGG} A. Cavagna, J.P. Garrahan and I. Giardina, Phys. Rev. B {\bf 61}, 
3960 (2000).


\bibitem{Douglas} M.R. Douglas, JHEP {\bf 05}, 046 (2003).

\bibitem{aazami} A. Aazami and R. Easther, JCAP03 p. 013 (2006).

\bibitem{Mersini-Houghton05}
L. Mersini-Houghton, Class. Quant. Grav. {\bf 22},  3481  (2005).

\bibitem{DM} D.S. Dean and S.N. Majumdar, Phys. Rev. Lett. {\bf 97}, 160201 
(2006); Phys. Rev. E {\bf 77}, 41108 (2008).

\bibitem{vivo2007large} P. Vivo, S.N. Majumdar and O. Bohigas, J. Phys. A: Math. Theor. {\bf 40}, 4317 (2007).

\bibitem{majumdar:060601} S.N. Majumdar and M. Vergassola, Phys. Rev. Lett. {\bf 102}, 060601 (2009).

\bibitem{nadal:061117} C. Nadal and S.N. Majumdar, Phys. Rev. E {\bf 79}, 061117 (2009).

\bibitem{vivo:216809} P. Vivo, S.N. Majumdar and O. Bohigas, Phys. Rev. Lett. {\bf 101}, 216809 (2008).

\bibitem{bray:150201} A.J. Bray and D.S. Dean, Phys. Rev. Lett. {\bf 98}, 150201 (2007).

\bibitem{fyod} Y.V. Fyodorov and I. Williams, J. Stat. Phys. {\bf 129}, 1081 (2007).

\bibitem{facchi:050502} P. Facchi, U. Marzolino, G. Parisi, S. Pascazio and A. Scardicchio, Phys. Rev. Lett. {\bf 101}, 050502 (2008).

\bibitem{kaz} P. Kazakopoulos, P. Mertikopoulos, A.L. Moustakas and G. Caire, [arXiv:0907.5024] (2009).

\bibitem{Tricomi} F.G. Tricomi, {\it Integral Equations} (Pure Appl. Math V, Interscience, London, 1957).

\bibitem{details} details will be published elsewhere.

\bibitem{chen2005orthogonal} Y. Chen and G. Pruessner, J. Phys. A: Math. Gen. {\bf 38}, L191 (2005).



\bibitem{di19952d} P. Di Francesco, P. Ginsparg and J. Zinn-Justin, Phys. Rep. {\bf 254}, 133 (1995).


















\end{thebibliography}

\end{document}